\documentstyle[12pt]{article}
\begin{document}
\begin{titlepage}
{\hskip 11cm} 
\vspace*{2.7cm}
\begin{center}
{\Large The Spectrum of the Hybrid Mesons with Heavy Quarks from 
the B.S. Equation\\}
\vspace*{0.4cm}
{{J.Y. Cui $^a$ $^ b$  H.Y. Jin $^a$ and J.M. Wu $^a$}\\
  {$^a$ \small Institute of High Energy Physics}\\
  {\small Beijing, 100039, P.R. China}\\
  {$^b$ \small Department of Physics, Henan Normal University}\\
  {\small Xinxiang, 453002, P.R. China}}
\end{center}
\date{}
\begin{center}
\begin{minipage}{120mm}
\vspace*{1.5cm}
\begin{center}{\bf Abstract}\end{center}
{We construct the B.S. equation for the hybrid mesons under instantaneous
approximation.  The kernel is chosen as the sum of an one-gluon exchange
potential and a linear confining
potential. The equations are solved by numerical method, and the spectrum of 
hybrid mesons $b\bar{b}g$ and $c\bar{c}g$ are obtained.\\
{\bf PACS: 14.40.C, 11.10.St. }}
\end{minipage}
\end{center}
\vskip 1in
\end{titlepage}
\newpage
1.{\it Introduction}${ ~~} $  Besides the conventional hadrons, quantum
chromodynamics, the theory of strong interaction, predicts that the states
containing gluonic excitations, namely glueballs and hybrid hadrons may exist.
The confirmation of the existence of such states would be a strong support
for QCD to be the true theory of strong interaction. Therefore, such states
attract a lot of interest both from theories and experiments. 
Several experimental candidates of glueballs and hybrid
states have been reported. For example, 
a candidate for the scalar glueball was reported by Crystal 
Barrel Collaboration at LEAR [1], and  possible evidence for a $1^{-+}$
light exotic hybrid candidate was reported by VES [2].  
As experiments begin to focus on the search for glueballs and hybrid states,
it's important to get more theoretical knowledge of such states from theory. In
this paper we will concentrate on the hybrid states with heavy quarks. 
\par
Many methods have been applied in the investigation of hybrid
 in the literature, which include the flux tube model [3], the bag model [4],
the constituent gluon model [5], and lattice Monte Carlo simulations [6].
In bag model, the mass of the lowest hybrid mesons with light quarks is about
$1.5GeV$. The hybrid mesons with heavy quarks are studied by taking into account
the bag deformation, and it's found that the masses of the lightest hybrid are
$3.9GeV$ for $c\bar{c}g$, and $10.5GeV$ for $b\bar{b}g$. In the flux tube model,
hybrid is interpreted as the phonon-type excitations of the string connecting
the quark and antiquark. It predicts that the masses of the lowest hybrid mesons
are $1.8-2.0GeV$ for $u\bar{u}g$ or $d\bar{d}g$, $4.2-4.5GeV$ for $c\bar{c}g$,
and $11.2GeV$ for $b \bar{b}g$.
\par The constituent gluon models for hybrid hadrons were introduced by Horn and
Madula [5]. In these models the constituent gluon plays the same role just as the
constituent quarks. It's assumed that the constituent gluon can have a
finite effective mass $m_g$ and carries angular momentum $l_g$. As a
result, it predicts non-exotic quantum numbers for the ground states of
hybrid. Exotic hybrid states including $1^{-+}$ and $0^{--}$ are predicted in the
higher-lying multiplet.
\par In this paper, we study the hybrid mesons wiht heavy quark
in the framework of B.S. equation.
This method is also based on the idea of constituent gluon. We only
concentrate on the hybrid with heavy quarks. In contrast to the light quark
sector, where the complicated flavor mixing between $q\bar{q}$ and $q\bar{q}g$
bound states may be important, and the $q\bar{q}$ spectrum 
itself is not very clear, for the 
heavy quark system, the $Q\bar{Q}$ spectrum is well described by the the
potential model, and the mixing between $Q\bar{Q}$ and $Q\bar{Q}g$ is small.
This implies that it is possible to find pure hybrid states in heavy
quark sector.
\par It's also interesting to note that in the constituent gluon model for
hybrid, when the mass of the quark and antiquark tends to infinity,
spin-symmetry would show up. We will make a more detailed discussion of
this point later.
\par The paper is arranged as follows. In section 2, we construct the B.S.
equation for the hybrid states. In section 3, we construct the B.S. wave
function of hybrid states. In section 4, the numerical result and a
discussion are presented.

2.{\it The B.S. equation of hybrid states}\hskip 0.2in  Let $\psi(x_1)$ and
$\bar{\psi}(x_2)$ be the quark and antiquark fields at points $x_1$ and $x_2$,
$A_\mu(x_3)$ the gluon field at point $x_3$. Then the B.S. wave function is
defined as

\begin{equation}
\chi_\mu (x_1,x_2,x_3)=\langle 0|T\psi(x_1)\bar{\psi}(x_2)A_\mu(x_3)|M
\rangle ,
\end {equation}
where $|M\rangle$ is the hybrid state with mass $M$ and momentum $P$. The color
indices has been suppressed. We define the relative and center of
mass kinematic variables as
\begin{equation}
X=\frac{\eta_1}{2}(x_1+x_2)+\eta_2x_3 ,
\end{equation}
\begin{equation}
x=x_1-x_2 ,\hskip 0.5in 
x^\prime=\frac{x_1+x_2}{2}-x_3 ,
\end{equation}
where $\displaystyle\eta_1=\frac{2m_Q}{2m_Q+m_g}, \eta_2=\frac{m_g}{2m_Q+m_g}$.\\
\vskip 0pt
With the translation invariance, we have
\begin{equation}
\chi_\mu(x_1,x_2,x_3)=e^{-iP\!\cdot\!X}\chi_\mu(x,x^\prime) .
\end{equation}
We can further define the B.S. wave function in momentum space by the
Fourier transformation
\begin{equation}
\chi_\mu(P,q,k)=\int d^4x d^4x^\prime e^{i(qx+k
x^\prime)}\chi_\mu(x,x^\prime) ,
\end{equation}
where $q$ is the relative momentum between the heavy quark and  heavy
antiquark, and $k$ is the relative momentum between the constituent gluon and
the center of the two heavy quarks. With a standard method we can obtain the
B.S. equation for the hybrid state.
\begin{equation}
\chi_\mu(P,q,k)=-\frac{1}{p\!\!\!/_1-m_Q}
\frac{i}{p_3^2-m_g^2}\int
\frac{d^4q^\prime}{(2\pi)^4}\frac{d^4k^\prime}{(2\pi)^4}
G_{\mu\nu}(P,q,q^\prime,k,k^\prime)
\chi^\nu(P,q^\prime,k^\prime)\frac{1}{p\!\!\!/_2+m_Q} .
\end{equation}
In obtaining the above equation, the propagators have been replaced by the
free ones.
$p_1$, $p_2$ and $p_3$ are the momentum of quark, antiquark and gluon
respectively, and we have the following relations
$$
\begin{array}{l}
\displaystyle p_1=\frac{\eta_1}{2}P+q+\frac{k}{2} ,
\end{array}$$
$$\begin{array}{l}
\displaystyle p_2=\frac{\eta_1}{2}P-q+\frac{k}{2} ,
\end{array}
\eqno(7)$$
$$\begin{array}{l}
\displaystyle p_3=\eta_2 P-k .  \hskip 0.4in
\end{array}$$
The integral kernel $G_{\mu\nu}$ in the above equation is defined as the sum of all
three-particle irreducible graphs. However, it's impossible to calculate the
kernel exactly from the first principle. In practice, some approximations and
phenomenological assumptions have to be made in obtaining the kernel. As usual,
we divide the kernel $G_{\mu\nu}$ into two parts: the long distance part
$G_{\mu\nu}^{(l)}$ and the short distance $G_{\mu\nu}^{(s)}$. $G_{\mu\nu}^{(l)}$
can be obtained approximately by calculating the following three
one-gluon-exchange diagrams shown in fig.1.
\par Besides equation (7), we also have
$$\begin{array}{l}
\displaystyle p_1^\prime=\frac{\eta_1}{2}P+q^\prime+\frac{k^\prime}{2} ,
\end{array}$$
$$\begin{array}{l}
\displaystyle p_2^\prime=\frac{\eta_1}{2}P-q^\prime+\frac{k^\prime}{2} ,
\end{array}
\eqno(8)$$
$$\begin{array}{l}
\displaystyle p_3^\prime=\eta_2 P-k^\prime . \hskip 0.5in
\end{array}$$
Hybrid is color singlet. Each of the two-body system in the hybrid belongs
to a definite representation of color SU(3). The $Qg$ system
belongs to a ${\bf 3}$, the $\bar{Q}g$ system belongs to a ${\bf 3}^*$, and
the $Q\bar{Q}$ system belongs to a ${\bf 8}$. The color factor for the
$Qg$ and $\bar{Q}g$ systems is $\displaystyle -\frac{3}{2}$, while the color
factor for $Q\bar{Q}$ system is $\displaystyle\frac{1}{6}$. 
This implies that the force between the 
constituent gluon and quark or antiquark is attractive, while the force
between the color-octet quark--antiquark pair is repulsive and much week.
Therefore, we have such a picture for the hybrid with heavy quarks: the
constituent quark and antiquark are almost static with weak and repulsive
force between them; the constituent gluon turns round the $Q\bar{Q}$ pair
and bind them together to form a bound state. Because the force between the 
quark 
and antiquark is very weak, in our treatment we simply neglect it. Then for
the short distance part of the kernel, we have (see Fig. 1)

$$\begin{array}{lcl}
G_{\mu\nu}^{(s)}(P,q,q^\prime,k,k^\prime)\!\!\!&= &
\!\!\! (2\pi)^4\delta^4(p_1-p_1^\prime)
\displaystyle\frac{3(4\pi\alpha_s)}{2~\!{\bf
l}^2}\Gamma_{\mu\nu\rho}\gamma^\rho(p\!\!/_1-m_Q)\\
\end{array}$$
$$\begin{array}{lcl}
\hskip 1.2in &+&\!\!\! \displaystyle (2\pi)^4\delta^4(p_2-p_2^\prime)
\frac{3(4\pi\alpha_s)}{2~\!{\bf l}^2}
\Gamma_{\mu\nu\rho}\gamma^\rho(p\!\!/_2+m_Q) ,
\end{array}
\eqno(9)$$
where $l=k-k^\prime$, and $\Gamma_{\mu\nu\rho}$ is the three-gluon-vertex,
$$
\begin{array}{c}
\Gamma_{\mu\nu\rho}=(p_3+p_3^\prime)_\rho g_{\mu\nu}+(l-p_3)_\nu g_{\mu\rho}
-(l+p_3^\prime)_\mu g_{\nu\rho} .
\end{array}
\eqno(10)$$
As for the long distance part of the kernel, we have to construct it
phenonemnologically. According to the experience with the ordinary mesons in
the potential model, the long distance part of the kernel is mainly of
scalar property. Therefore we only choose the first term of equation (10) as
the spin dependence of the long distance part of the kernel, and we assume
$$\begin{array}{l}
\displaystyle G_{\mu\nu}^{(l)}=-g_{\mu\nu}(p_3+p_3^\prime)\!\cdot\!V
\frac{8\pi\lambda}{{\bf l}^4}\left(\frac{}{}(2\pi)^4\delta^4(p_1-p_1^\prime)
(p\!\!/_1-m_Q) - (2\pi)^4\delta^4(p_2-p_2^\prime)(p\!\!/_2+m_Q)\right).
\end{array}
\eqno(11)$$
where $\displaystyle V=\frac{P}{M}$, and  
$\displaystyle\frac{8\pi \lambda}{{\bf l}^4}$ corresponds to a linear 
potential in the position space.
\par Substituting the kernel into equation (6), and complete the integration
over $q^\prime$, we obtain
$$\begin{array}{l}
\chi_\mu(P,k)\! = \displaystyle \frac{-i}{p_3^2-m_g^2}
\int\frac{d^4k^\prime}{(2\pi)^4}
\!\left(\frac{3(4\pi\alpha_s)}{2~\!{\bf
l}^2}\Gamma_{\mu\nu\rho}\gamma^\rho - g_{\mu\nu}(p_3+p_3^\prime)\!\cdot\!V
\frac{8\pi\lambda}{{\bf
l}^4}\right)\! \chi^\nu(P,k^\prime)\frac{1}{p\!\!/_2+m_Q}\\
~\\
\hskip 0.6in  -  \displaystyle\frac{i}{p_3^2-m_g^2}\frac{1}{p\!\!/_1-m_Q}
\int\frac{d^4k^\prime}{(2\pi)^4} \left( \frac{3(4\pi\alpha_s)}{2~\!{\bf
l}^2}\Gamma_{\mu\nu\rho}\gamma^\rho + g_{\mu\nu}(p_3+p_3^\prime)\!\cdot\!V
\frac{8\pi\lambda}{{\bf l}^4}\right) \chi^\nu(P,k^\prime). 
\end{array}
\eqno(12)$$
We see that now $q$ is not a dynamical variable. By neglecting the $q$
dependence of the propagators, this variable is integrated out.
The propagator of the quark can be expressed as
$$\begin{array}{l}
\displaystyle\frac{1}{p\!\!/_1-m_Q}=(\frac{\Lambda^+({\bf p}_1)}{p_{10}-E_1}+
\frac{\Lambda^-({\bf p}_1)}{p_{10}+E_1})\gamma_0 ,
\end{array}
\eqno(13)$$
$$\begin{array}{l}
\displaystyle\frac{1}{p\!\!/_2+m_Q}=
\gamma_0(\frac{\Lambda^+({\bf p}_2)}{p_{20}+E_2}+
\frac{\Lambda^-({\bf p}_2)}{p_{20}-E_2}) ,
\end{array}
\eqno(14)$$
where $E_i=\sqrt{m_i^2+{\bf p}_i^2}$, and
$\Lambda^+({\bf p})$ $(\Lambda^-({\bf p}))$ is positive (negative) energy
projector defined as
$$\begin{array}{l}
\displaystyle\Lambda^{\pm}({\bf p}_i)=\frac{E_i\pm\gamma_0
({\bf \gamma}\!\cdot\!{\bf p}_i +m_Q)}{2E_i} .
\end{array}
\eqno(15)$$
For the constituent gluon propagator, we have
$$\begin{array}{l}
\displaystyle\frac{1}{p_3^2-m_g^2}=
\frac{1}{2E_3}(\frac{1}{p_{30}-E_3}-\frac{1}{p_{30}+E_3}) .
\end{array}
\eqno(16)$$
As an approximation, we neglect the negative energy part of the propagator.
Such approximation is reasonable especially for heavy particles.
Then as usual, instantaneous approximation is made.  Under this
approximation we ignore the $dk_0$ dependence of the kernel. After making
$k_0$ integration over the two sides of equation (12), we have
$$\begin{array}{l}
2E_3(M- E_1- E_2-E_3)\phi_\mu(k)\\
\end{array}$$
$$\begin{array}{l}
 = -
\displaystyle \int\frac{d^3k^\prime}
{(2\pi)^3}\left( \frac{3(4\pi\alpha_s)}{2~{\bf
l}^2}\Gamma_{\mu\nu\rho}\gamma^\rho - g_{\mu\nu}(p_3+p_3^\prime)\!\cdot\!V
\frac{8\pi\lambda}{{\bf l}^4} \right) 
\phi^\nu(k^\prime)\gamma_0\Lambda^-({\bf p}_2)\\
\end{array}$$
$$\begin{array}{l}
\hskip 0.5in - ~ \displaystyle\Lambda^+({\bf p}_1)
\gamma_0\int\frac{d^3k^\prime}
{(2\pi)^3} \left( \frac{3(4\pi\alpha_s)}{2~{\bf
l}^2}\Gamma_{\mu\nu\rho}\gamma^\rho + g_{\mu\nu}(p_3+p_3^\prime)\!\cdot\!V
\frac{8\pi\lambda}{{\bf l}^4}\right) \phi^\nu(k^\prime) ,
\end{array}\eqno(17)$$
where $\phi_\mu(k)$ is the instantaneous three dimensional B.S. wave function.
$$ \begin{array}{l}
\displaystyle\phi_\mu(P,k)=\int dk_0\chi_\mu(P,k)
\end{array}
\eqno(18)$$
\par
3. {\it The B.S. wave function of the hybrid}{~~~~} In order to solve the B.S.
equation, we should construct the B.S. wave function of the hybrid with the
given quantum numbers. In the following we construct the wave function in
heavy quark limit.
The spin of the quark-antiquark system, $S_{Q\bar{Q}}$, can take two
values, $0$ and $1$. When $S_{Q\bar{Q}}=0$, we have the following hybrid
states and its four dimensional wave functions.\\
The $1^{+-}$ hybrid:
$$
\begin{array}{l}
\chi_\mu(P,k)=(1+V\!\!\!\!\!/~)\gamma_5(F_1e_\mu + F_2k\!\cdot\!e ~ k_{\bot\mu}
+F_3k\!\cdot\!eP_\mu).
\end{array}
\eqno(19)$$
The $0^{--}$ hybrid
$$ \begin{array}{l}
\chi_\mu(P,k)=(1+V\!\!\!\!\!/~)\gamma_5(F_1k_{\perp\mu} + F_2P_\mu).
\end{array}
\eqno(20)$$
The $1^{--}$ hybrid
$$\begin{array}{l}
\chi_\mu(P,k)=F(1+V\!\!\!\!\!/~)\gamma_5\varepsilon_{\mu\nu\rho\sigma}P^\nu
k_\bot^\rho e^\sigma .
\end{array}
\eqno(21)$$
The $2^{--}$ hybrid
$$\begin{array}{l}
\chi_\mu(P,k)=(1+V\!\!\!\!\!/~)\gamma_5(F_1\eta_{\mu\nu}k^\nu
+F_2\eta_{\lambda\nu}k^\lambda k^\nu
k_{\bot\mu}+F_3\eta_{\lambda\nu}k^\lambda k^\nu P_\mu).
\end{array}
\eqno(22)$$
\ldots\ldots 

\par When $S_{Q\bar{Q}}=1$ we have another series of hybrid states and its four
dimensional wave functions.\\
The $0^{++}$ hybrid
$$\begin{array}{l}
\chi_\mu(P,k)=(1+V\!\!\!\!\!/~)(F_1\gamma_\mu + F_2k\!\!\!/_\bot k_{\bot\mu}
+ F_3k\!\!\!/_\bot P_\mu).
\end{array}
\eqno(23)$$
The $1^{++}$ hybrid
$$\begin{array}{l}
\chi_\mu(P,k)  =
(1+V\!\!\!\!\!/~)(F_1\varepsilon_{\mu\nu\rho\sigma}\gamma^\nu e^\rho
P^\sigma + F_2\varepsilon_{\mu\nu\rho\sigma}\gamma_\bot^\nu e^\rho k_\bot^\sigma
+ F_3\varepsilon_{\mu\nu\rho\sigma}e^\nu P^\rho k^\sigma k\!\!\!/_\bot
\hskip 0.8in
\end{array}$$
$$\begin{array}{l}
+F_4\varepsilon_{\mu\nu\rho\sigma}\gamma^\nu P^\rho k^\sigma k\!\cdot\!e
+F_5\varepsilon_{\lambda\nu\rho\sigma}\gamma^\lambda e^\nu P^\rho k^\sigma P_\mu
+F_6\varepsilon_{\lambda\nu\rho\sigma}\gamma^\lambda e^\nu P^\rho k^\sigma
k_{\bot\mu}).
\end{array}
\eqno(24)$$
The $2^{++}$ hybrid
$$\begin{array}{l}
\chi_\mu(P,k)=
(1+V\!\!\!\!\!/~)(F_1\eta_{\mu\nu}\gamma^\nu
+F_2\eta_{\mu\nu}k^\nu k\!\!\!/_\bot
+F_3\eta_{\rho\sigma}\gamma^\rho k^\sigma P_\mu . \hskip 1.9in
\end{array}$$
$$\begin{array}{l}
+ F_4\eta_{\rho\sigma}\gamma^\rho k^\sigma k_{\bot\mu}
+ F_5\eta_{\rho\sigma}k^\rho k^\sigma \gamma_{\bot\mu}
+ F_6\eta_{\rho\sigma}k^\rho k^\sigma k\!\!\!/_\bot P_\mu
+ F_7\eta_{\rho\sigma}k^\rho k^\sigma k\!\!\!/_\bot k_{\bot\mu}).
\end{array}
\eqno(25)$$
The $0^{-+}$ hybrid
$$\begin{array}{l}
\chi_\mu(P,k)=F(1+V\!\!\!\!\!/~)\varepsilon_{\mu\nu\rho\sigma}\gamma^\nu P^\rho
k^\sigma .
\end{array}
\eqno(26)$$
The $1^{-+}$ hybrid
$$\begin{array}{l}
\chi_\mu(P,k)=(1+V\!\!\!\!\!/~)(F_1e\!\!/k_{\bot\mu}
+ F_2e\!\cdot\!k\gamma_{\bot\mu}
+ F_3k\!\!\!/_\bot e_\mu
+ F_4e\!\!\!/P_\mu + F_5k\!\!\!/_\bot k\!\cdot\!ek_{\bot\mu}
+ F_6k\!\!\!/_\bot k\!\cdot\!eP_\mu).
\end{array}
\eqno(27)$$
The $2^{-+}$ hybrid
$$\begin{array}{l}
\chi_\mu(P,k)=
(1+V\!\!\!\!\!/~)(F_1\varepsilon_{\mu\nu\rho\sigma}\eta^{\nu\lambda}
P^\rho\gamma^\sigma k_\lambda
+ F_2\varepsilon_{\mu\nu\rho\sigma}\eta^{\nu\lambda}
P^\rho k^\sigma\gamma_\lambda + F_3\varepsilon_{\mu\nu\rho\sigma}
\eta^{\nu\lambda}\gamma_\bot^\rho k_\bot^\sigma k_\lambda . \hskip 0.4in
\end{array}$$
$$\begin{array}{l}
+ F_4\varepsilon_{\mu\nu\rho\sigma}
\eta^{\nu\lambda}P^\rho k^\sigma k_\lambda k\!\!\!/_\bot
+ F_5\varepsilon_{\lambda\nu\rho\sigma}\eta_\mu^\lambda P^\nu\gamma^\rho k^\sigma
+ F_6\varepsilon_{\lambda\nu\rho\sigma}\eta^{\lambda\alpha}P^\nu\gamma^\rho
k^\sigma k_\alpha k_{\bot\mu}. 
\end{array}$$
$$\begin{array}{l}
+ F_7\varepsilon_{\lambda\nu\rho\sigma}
\eta^{\lambda\alpha}P^\nu\gamma^\rho k^\sigma k_\alpha P_\mu). \hskip 3.0in
\end{array}
\eqno(28)$$
\ldots\ldots\\
In the above expressions $e_\mu$ is the polarization vector of the vector hybrid, and $\eta_{\mu\nu}$
the polarization tensor of the tensor hybrid. $q_{\bot \mu}$ is a transverse vector
defined as $q_{\bot\mu}=q_\mu - q\!\cdot\! VV_\mu$, and $\gamma_{\bot\mu}$ is
defined similarly. $F_i$'s are scalar functions of $q^2$, $P\!\cdot\! q$. 
For simlicity, the same $F_i$ in different hybrid states represent different 
functions.
\par
If we substitute the wave function into the B.S. equation, the spin symmetry
can be seen explicitly. For example, for $1^{++}$ state, the components of
 its wave function, $F_1$, $F_5$ and $F_6$ are decoupled from the total wave
function $\chi_{\mu}$, and the equations about these components are exactly
the same as those of $0^{++}$ hybrid. Similarly, for $2^{++}$ state, the
components of the wave function, $F_1$, $F_3$ and $F_4$ are decoupled out
from the total wave function, and the equations about these components are
also the same as those of $0^{++}$ hybrid. The reason is that when the
interaction between $Q\bar{Q}$ is week and negligible, the quark spin
decouples from the light freedom in heavy quark limit. This result does not
depend on the form of the kernel (10) and valid for more general case in
which the interaction between $Q\bar{Q}$ is omitted. 
Therefore, if the constituent
gluon model is true, we wish to see this symmetry at least in the low-lying
$Q\bar{Q}g$ hybrid.  

4.{\it The numerical results}\hskip 0.2in 
In the non-relativistic limit, in our numerical treatment we only choose
the spatial part of the wave function. In the center of mass frame, the form
of the wave functions are simplified significantly.\\
For $1^{+-}$ state
$$\begin{array}{l}
\phi_i(k)=(1+\gamma_0)\gamma_5 (f_1e_i+f_2 e\!\cdot\! k k_i).
\end{array}
\eqno(29)$$
For $0^{--}$ state
$$\begin{array}{l}
\phi_i(k)=f_1(1+\gamma_0)\gamma_5k_i .
\end{array}
\eqno(30)$$
For $1^{--}$ state
$$\begin{array}{l}
\phi_i(k)=f_1(1+\gamma_0)\gamma_5\varepsilon_{ijk}k_je_k .
\end{array}
\eqno(31)$$
For $2^{--}$ state
$$\begin{array}{l}
\phi_i(k)=(1+\gamma_0)\gamma_5(f_1\eta_{ij}k_j+f_2\eta_{jk}k_j k_k k_i).
\end{array}
\eqno(32)$$
For $0^{++}$ state
$$\begin{array}{l}
\phi_i(k)=(1+\gamma_0)(f_1\gamma_i+f_2\gamma\!\cdot\!k~k_i).
\end{array}
\eqno(33)$$
For $1^{++}$ state
$$\begin{array}{l}
\phi_i(k)=(1+\gamma_0)(f_1\varepsilon_{ijk}\gamma_je_k+\varepsilon_{ijk}
k_je_k k\!\cdot\!\gamma +f_3\varepsilon_{ijk}k_j\gamma_k k\!\cdot\!e
+f_4\varepsilon_{jkl}k_je_k\gamma_l k_i).
\end{array}
\eqno(34)$$
For $2^{++}$ state
$$\begin{array}{l}
\phi_i(k)=(1+\gamma_0)(f_1\eta_{ij}\gamma_j+f_2\eta_{ij}k_j k\!\cdot\!\gamma
+f_3\eta_{jk}k_j\gamma_kg_i+f_4\eta_{jk}k_j k_k\gamma_i
+f_5\eta_{jk}k_j k_k k\!\cdot\!\gamma k_i).
\end{array}
\eqno(35)$$
\ldots\ldots 
\par
Substituting the above wave functions into the B.S. equation, we can calculate
the spectrum of hybrid states. It's easy to see that the states $1^{+-}$ and
$(0,1,2)^{++}$ degenerate, and also the states $(0,1,2)^{--}$ and
$(0,1,2)^{-+}$ degenerate.
\par
It's easy to see that the long distance part of the kernel with the form of 
expression (11) is very 
singular at the point ${\bf l}^2= 0$. Some form of regularization is necessary.
The method is to make the following replacement
$$\begin{array}{l}
\displaystyle G_{\mu\nu}^{(l)}\!\rightarrow\! -g_{\mu\nu}
(p_3\!+\!p_3^\prime)\!\cdot\!V\!
\left(\frac{}{}\!(2\pi)^4\delta^4(p_1\!-\!p_1^\prime)
(p\!\!/_1 \!-\! m_Q)\! - \!(2\pi)^4\delta^4(p_2\! -\! p_2^\prime)
(p\!\!/_2+m_Q)\!\right)
\frac{8\pi\lambda}{({\bf l}^2+u^2)^2}   
\end{array}$$
$$\begin{array}{l}
 +  \delta^3({\bf l})\displaystyle\int d^3{\bf k}\left\{g_{\mu\nu}
 (p_3+p_3^\prime)\!\cdot\!V \frac{8\pi\lambda}{({\bf l}^2+u^2)^2}
\left(\frac{}{}(2\pi)^4\delta^4(p_1-p_1^\prime)
(p\!\!/_1-m_Q)
\right.\right.
\end{array}$$
$$\begin{array}{l}
\displaystyle\left.\left.
-(2\pi)^4\delta^4(p_2-p_2^\prime)(p\!\!/_2+m_Q)\frac{}{}\right)\right\}
\end{array}
\eqno(36)$$
where $u$ is a small quantity. In actual calculation, we let $u\rightarrow 0$. 
 In this way, the infared divergence is subtracted out.
\par In our calculation, the strong coupling constant $\alpha_s$ is chosen
as a running one.
$$\begin{array}{l}
ð\displaystyle\alpha_s=\frac{12\pi}{27}\frac{1}{ln(a+l^2/\Lambda_{QCD}^2)}
\end{array}
\eqno(37)$$
where the parameter $a$ is introduced to avoid the infrared divergence.
We choose $a=4.0$ which implies that $\alpha_s$ tends to its largest value $1$
when $l^2\rightarrow 0$.
\par
The other parameters including the string tension $\lambda$, and the quark
mass, $m_c$ and $m_b$, are determined by fitting the $Q\bar{Q}$ spectrum, and
we have: $\lambda=0.2(GeV)^2$, $m_c=1.47GeV$ and $m_b=4.80GeV$.
\par The mass of the constituent gluon has been studied in a lattice
calculation [7], and in non-perturbative QCD based on gauge invariant
Lagrangian [8]. These studies yield $m_g=(500\sim 800)MeV$. In our study, we
adopt $m_g=600MeV$.
\par   All the parameters are determined as above, we can solve the B.S.
equation numerically. The results are list in the table 1.
\par As mentioned above, the choice $m_g=600MeV$ is reasonable. However, we
can't say it's exact. We calculate the hybrid mass by changing $m_g$ in
the the range of several hundred $MeV$. Fig. 2 shows the dependence of the
mass of $c\bar{c}g$ on $m_g$, and fig.3 shows the dependence of the mass of
$b\bar{b}g$ on $m_g$. We can see that the hybrid mass is not too sensitive
to the gluon mass. The mass of lightest $c\bar{c}g$ changes from $3.65GeV$ to
$3.80GeV$ when $m_g$ changes from $300MeV$ to $800MeV$. In the same range of
$m_g$, the mass of the lightest $b\bar{b}g$ changes from $10.26GeV$ to
$10.39GeV$. 
\par There is another kind of uncertainty due to the neglecting of the
interaction between the quark and anti-quark. 
Recall that for the ground state of charmonium or
bottumonium in quark model, the interaction energy between the two quarks is
less than one hundred $MeV$, and here the strength of the interaction between
the color-octet $Q\bar{Q}$ is only one  eighth of that between the singlet
$Q\bar{Q}$. Therefore, we expect this uncertainty is small and can be
neglected reasonably.  
\par 5.~{\it Conclusion and discussion}{~~~~} We have studied the hybrid states
with heavy quarks in the framework of B.S. equation. In the heavy quark
limit, we show clearly the existence of the spin symmetry which results in the
degeneracy of the states of $(0,1,2)^{++}$ and also the degeneracy of
$(0,1,2)^{-+}$. The states $1^{+-}$ and $(0,1,2)^{++}$ are also degenerate,
and they are the ground state of hybrid, while the degenerate states
$(0,1,2)^{--}$ and $(0,1,2)^{-+}$ belong to the first excited states.
With a reasonable choice of
$m_g(600MeV)$, we find that the ground state of $c\bar{c}g$ hybrid is near
 the threshold of $D\bar{D}$ meons, while the ground state of $b\bar{b}g$
hybrid is bellow the threshold of $B\bar{B}$ mesons. The energy gap between the
first excited states and the ground states are $0.52GeV$ for $c\bar{c}g$, and
$0.48GeV$ for $b\bar{b}g$, and such values are quite insensitive to the mass
of the constituent gluon.
\par In the experiment of $e^+e^-$ collision, several resonant states are
found near $4.0GeV$, such as $\psi(4040)$, $\psi(4160)$ and $\psi(4415)$.
Within the conventional picture of charmonium, $\psi(4040)$ is assigned 
as $3S$ state, 
and $\psi(4415)$ is assigned as $4S$ (or possibly $5S$) state. But as has
been argued by F. Close [9], $\psi(4160)$ perhaps 
is a $1^{--}$ $c\bar{c}g$ 
hybrid state. In our model, the mass of the $1^{--}$ $c\bar{c}g$ hybrid is
$4.24GeV$, so our result is roughly in agreement with this interpretation.

\newpage

Figuer Caption\\
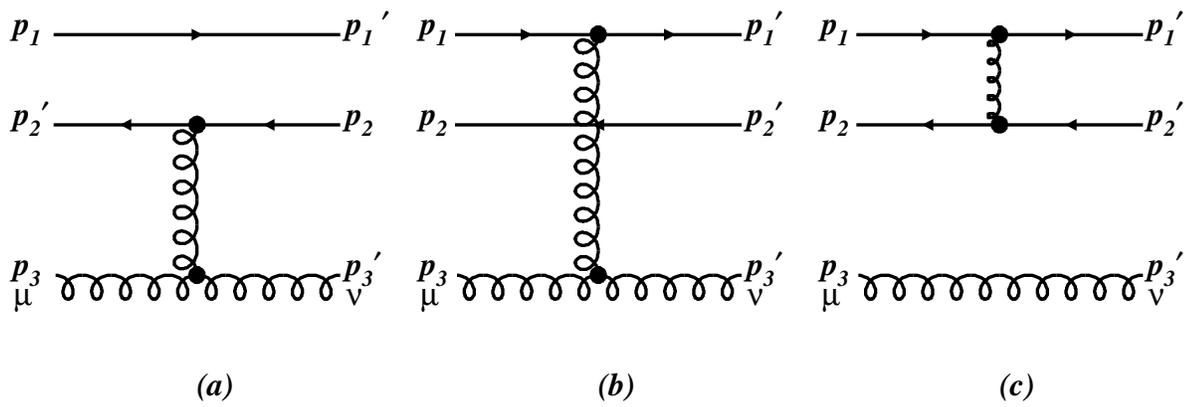
Fig.1:The diagrams of B.S. kernel in the lowest order.  \\
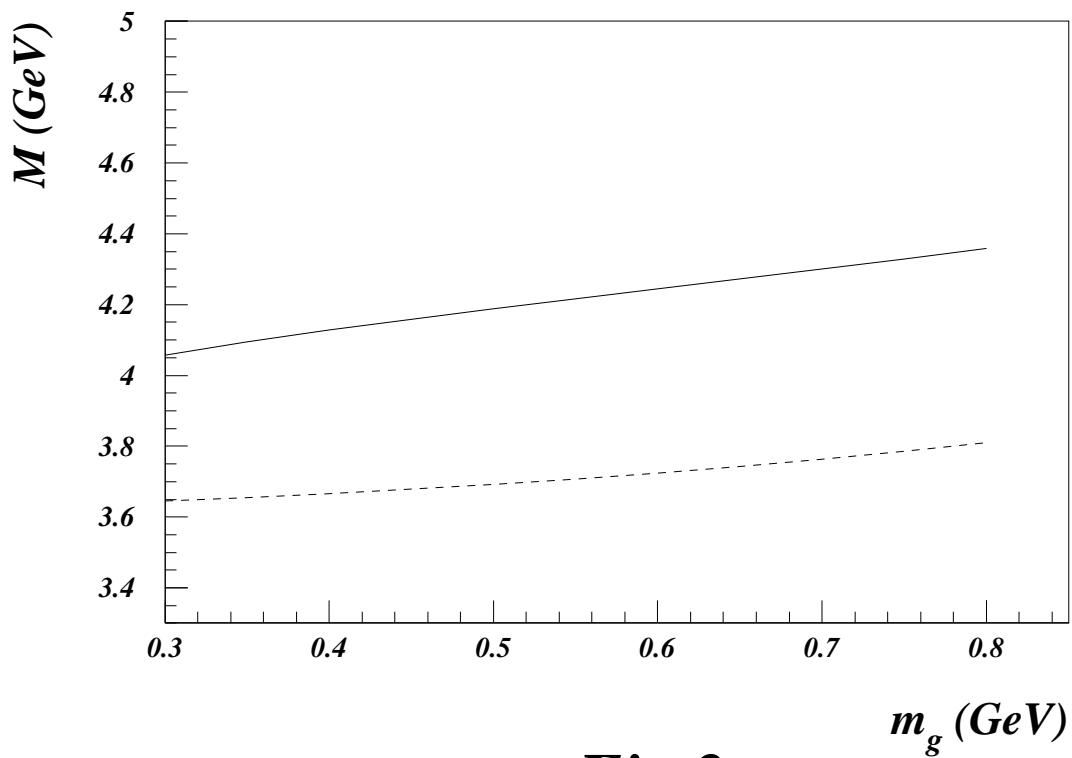
Fig.2:The $m_g$ dependence of the mass of the ground and
first excited $c\bar{c}g$ hybrid. \\
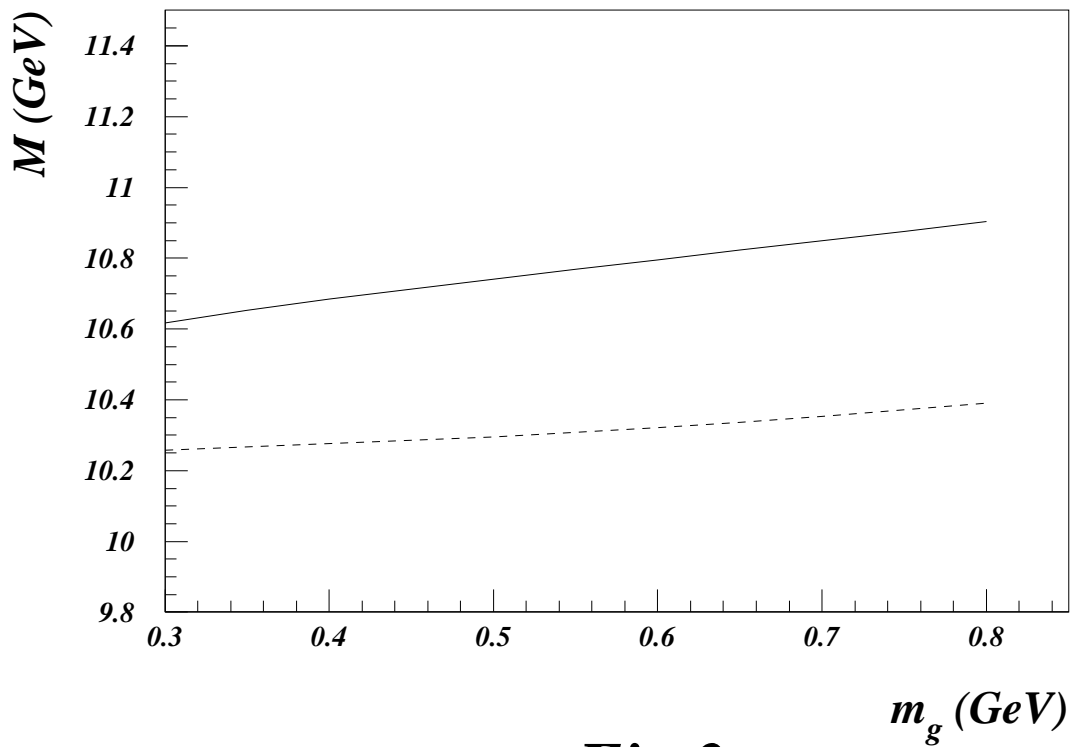
Fig.3: The $m_g$ dependence of the mass of the ground and
first excited $b\bar{b}g$ hybrid.\\

\newpage
Table Caption\\
Table 1: The mass of the hybrid mesons $(MeV)$. 
\end{document}